\documentclass[twocolumn,showpacs,preprintnumbers,amsmath,amssymb,superscriptaddress,groupedaddress,prmaterials]{revtex4}

\pdfoutput = 1
                  
\usepackage{epsfig}                                                            
\usepackage{graphicx}
\usepackage{dcolumn}
\usepackage{color}
\usepackage{bm}
\usepackage{subfigure}

\begin{document}

\title{\bf Origin of Ferroelectricity in Orthorhombic LuFeO$_3$}

\author {Ujjal Chowdhury} \affiliation  {Nanostructured Materials Division, CSIR-Central Glass and Ceramic Research Institute, Kolkata 700032, India}
\author {Sudipta Goswami} \affiliation {Department of Solid State Physics, Indian Association for the Cultivation of Science, Kolkata 700032, India }
\author {Amritendu Roy} \affiliation {School of Minerals, Metallurgical and Materials Engineering, Indian Institute of Technology, Bhubaneswar 752050, India}  
\author {S.S. Rajput} \affiliation {Department of Materials Science and Engineering, Indian Institute of Technology, Kanpur 208016, India}
\author {A.K. Mall} \affiliation {Materials Science Programme, Indian Institute of Technology, Kanpur 208016, India}
\author {R. Gupta} \affiliation {Department of Physics, Indian Institute of Technology, Kanpur 208016, India}
\author {S.D. Kaushik} \affiliation {UGC-DAE Consortium for Scientific Research, Bhabha Atomic Research Centre, Mumbai 400085, India}
\author {V. Siruguri} \affiliation  {UGC-DAE Consortium for Scientific Research, Bhabha Atomic Research Centre, Mumbai 400085, India}
\author {S. Saravanakumar} \affiliation {Department of Physics, Kalasalingam University, Krishnakoil 626126, India}
\author {S. Israel} \affiliation {Department of Physics, The American College, Madurai, India}
\author {R. Saravanan} \affiliation {Department of Physics, The Madura College, Madurai 625011, India}
\author {A. Senyshyn} \affiliation {Forschungsneutronenquelle Heinz Maier-Leibnitz (FRM II), Technische Universitat Munchen, D-85747 Garching b. Munchen, Germany}  
\author {T. Chatterji} \affiliation {Science Division, Institut Laue-Langevin, BP 156, 38042 Grenoble Cedex 9, France}
\author {J.F. Scott} \affiliation {School of Chemistry, University of St Andrews, St Andrews, Fife, KY16 9ST, United Kingdom} 
\author {A. Garg} \email {ashishg@iitk.ac.in} \affiliation {Department of Materials Science and Engineering, Indian Institute of Technology, Kanpur 208016, India} 
\author {Dipten Bhattacharya} 
\email{dipten@cgcri.res.in} \affiliation {Nanostructured Materials Division, CSIR-Central Glass and Ceramic Research Institute, Kolkata 700032, India}

\date{\today}

\begin{abstract}
We demonstrate that small but finite ferroelectric polarization ($\sim$0.01 $\mu$C/cm$^2$) emerges in orthorhombic LuFeO$_3$ ($Pnma$) at $T_N$ ($\sim$600 K) because of commensurate (k = 0) and collinear magnetic structure. The synchrotron x-ray and neutron diffraction data suggest that the polarization could originate from enhanced bond covalency together with subtle contribution from lattice. The theoretical calculations indicate enhancement of bond covalency as well as the possibility of structural transition to the polar $Pna2_1$ phase below $T_N$. The $Pna2_1$ phase, in fact, is found to be energetically favorable below $T_N$ in orthorhombic LuFeO$_3$ ($albeit$ with very small energy difference) than in isostructural and nonferroelectric LaFeO$_3$ or NdFeO$_3$. Application of electric field induces finite piezostriction in LuFeO$_3$ via electrostriction resulting in clear domain contrast images in piezoresponse force microscopy.            
\end{abstract} 
\maketitle

\section{Introduction}
During the last few years, work on ferroelectricity in rare-earth orthoferrites RFeO$_3$ (R = Sm, Dy, Tb, Y, Lu) poses quite a few puzzles. In SmFeO$_3$, YFeO$_3$, and LuFeO$_3$ \cite{Lee-1, Shang, Chowdhury-1}, the ferroelectric order is reported to set in right at the antiferromagnetic transition temperature $T_N$ ($\sim$600 K). On the other hand, in DyFeO$_3$ \cite{Tokunaga}, the ferroelectric transition takes place at a much lower temperature ($T_N^{Dy}$ $\sim$4 K) only when application of magnetic field $H$ $\parallel$ $c$ induces a ferromagnetic component to the Fe sublattice. While observation of ferroelectricity in DyFeO$_3$ still remains unchallenged, the ferroelectricity in SmFeO$_3$ below $T_N$ has been disputed from direct electrical measurement of polarization and capacitance-voltage characteristics \cite{Kuo} as well as from crystallography \cite{Johnson}. It has been pointed out that the rare-earth orthoferrites, in general, are paraelectric down to T = 0 and could only exhibit ferroelectricity in thin film form upon introduction of appropriate lattice strain \cite{Zhao}. If ferroelectricity at all emerges at $T_N$ in the bulk form of the sample, it should be due to the inversion-symmetry-breaking magnetic structure. The structure could either be noncollinear arising from antisymmetric Dzyloshinskii-Moriya or p-d exchange or collinear arising from exchange striction \cite{Tokura}. The collinear magnetic structure in SmFeO$_3$ seems to yield nonpolar $Pbnm$ although possibility of polar $m2m$ point group was also hinted \cite{Scott}. The controversy surrounding the emergence of ferroelectricity in orthorhombic SmFeO$_3$, therefore, calls for a thorough examination of the issue in other such rare-earth orthoferrites. Observation of finite ferroelectricity at $T_N$ in this class of compounds has got another important implication. If ferroelectricity is indeed observed in them at $T_N$, they can form a new class of room temperature Type-II multiferroics. 

\begin{figure}[!ht]
  \begin{center}
    \includegraphics[scale=0.30]{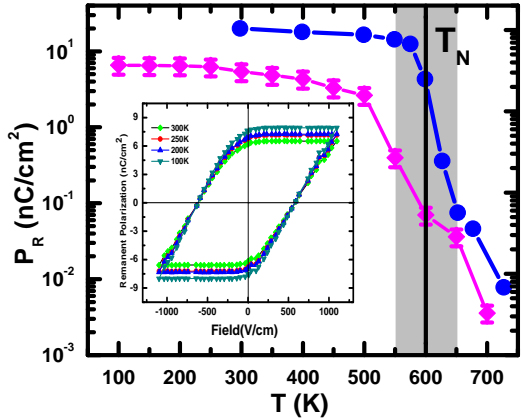} 
    \end{center}
  \caption{(color online) The variation of the remanent polarization (diamond) with temperature measured in orthorhombic LuFeO$_3$; the $P_{el}$ (circle) calculated from x-ray diffraction are also shown; inset shows the remanent hysteresis loops measured at different temperatures. }
\end{figure}

\begin{figure*}[!ht]
  \begin{center}
    \includegraphics[scale=0.70]{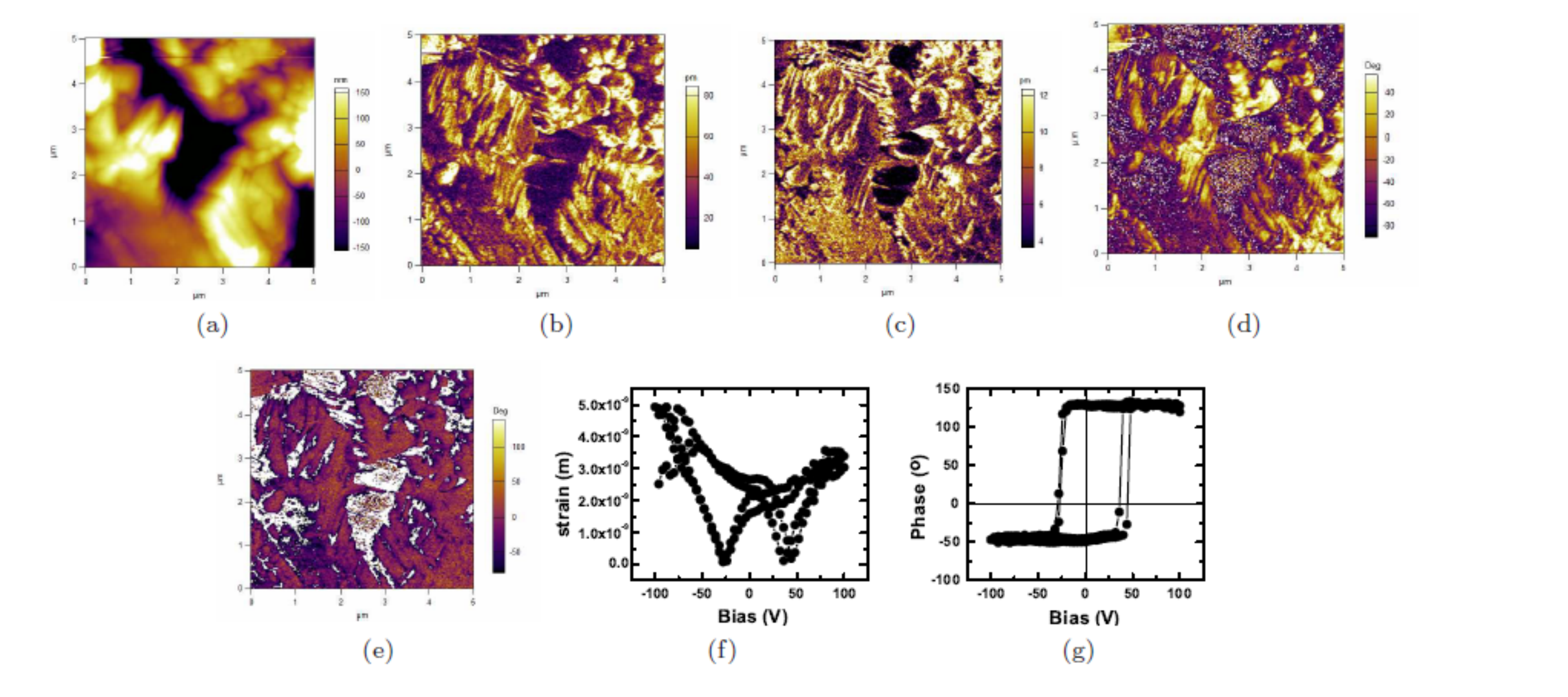} 
    \end{center}
  \caption{(color online) The (a) AFM and (b), (c) amplitude contrast and (d), (e) phase contrast PFM images under +100V and -100V bias voltage, respectively; (f) strain-field and (g) phase switching hysteresis loops. }
\end{figure*}

In this work, we examined the occurrence of ferroelectricity in orthorhombic LuFeO$_3$ at $T_N$($\sim$600 K). We employed a special protocol within the modified Sawyer-Tower circuit to extract the intrinsic remanent ferroelectric polarization. This is complemented by piezoresponse force microscopy. We investigated the electronic, crystallographic, and magnetic structures in the material using synchrotron x-ray and high resolution powder neutron diffraction experiments while Raman spectrometry was employed to track the phonon modes across the ferroelectric transition. The results collectively suggest that the tiny ferroelectric polarization ($\sim$0.01 $\mu$C/cm$^2$), emerging at the magnetic transition in orthorhombic LuFeO$_3$, could originate from enhanced bond covalency though subtle role of underlying lattice cannot be ruled out. The theoretical calculations conducted to investigate the origin of ferroelectricity show the contribution of both electronic and lattice structures to the observed polarization. In addition, possibility of a structural transition at $T_N$ (from $Pnma$ $\rightarrow$ $Pna2_1$) could also be observed which, however, because of small energy difference between these two phases (smaller than the room temperature thermal energy), could not be detected experimentally.  

\section{Experiments}
Experiments have been carried out on phase pure high quality bulk polycrystalline samples. The details of the sample preparation have been given elsewhere \cite{Chowdhury-1}. The synchrotron x-ray data were recorded at the MCX beamline of Elettra, Trieste ($\lambda$ = 0.61992 \AA) and the neutron diffraction experiments were carried out at the SPODI FRM-II diffractometer of Technische Universitat Munchen, Munchen ($\lambda$ = 2.536 \AA), and also at the PD-3 diffractometer of NFNBR, Bhabha Atomic Research Centre, Mumbai ($\lambda$ = 1.48 \AA). The synchrotron x-ray data have been refined by JANA 2006 and the structure factors were used to construct the charge density distribution map within a unit cell by employing Maximum Entropy Method (MEM). The neutron data have been refined by FullProf for determining the magnetic and crystallographic lattices. The remanent ferroelectric hysteresis loops were measured by PC Loop Tracer of Radiant Inc., and the piezoresponse force microscopy (PFM) were carried out by MFP-3D scanning probe microscope of Asylum. In addition, Raman spectra have been recorded across 300-700 K in order to track the phonon modes around the ferroelectric transition.

\section{Computational Details}
The first-principles calculations were performed using density functional theory as implemented in Vienna ab-initio Simulation Package (VASP). The generalized gradient approximations (GGA) by Perdew-Bruke and Ernzerhof \cite{Perdew} (PBEsol), optimized for solids, has been used. In order to verify the robustness, some of the calculations were tested using a different functional, PW-91 \cite{Wang-2}. The strongly correlated electrons of the transition metal ions within the optimized structure have been taken care of by the Hubbard potential ($U_{eff}$ = 3-5 eV) within (GGA+U) for a separate set of calculations wherein a rotationally invariant approach by Dudarev $\textit{et al}$. \cite{Dudarev} was adapted. We used projected augmented wave potentials and considered 9 valence electrons for Lu, 14 for Fe (including the semi-core states) and 6 for O ions. We used a Monkhorst-pack $k$-mesh of size 3$\times$4$\times$6 for all our calculations. A paramagnetic state is the outcome of counteracting long-range ordering of magnetic moments and thermal energy to disrupt the above ordering. Therefore, it can be presumed that at zero temperature (no thermal energy), the long-range ordering of the spins would assume the configuration that minimizes the total energy of the system. In view of the above, a 2$\times$1$\times$1 supercell was constructed and different spin configurations [ferromagnetic, A-antiferromagnetic (AFM), C-AFM, E-AFM and G-AFM] were enforced. Total energy corresponding to the above spin configurations was computed to determine the most favored magnetic ordering. To explore the possibility of a structural phase transition in presence of antiferromagnetic order, the experimentally observed structure with $Pnma$ symmetry at 298 K was transformed to $Pna2_1$ (one of the subgroups of $Pnma$) using TRANSTRU within the Bilbao Crystallographic server. The supercells with magnetic ordering and assuming either $Pnma$, distorted $Pnma$ or $Pna2_1$ structure were fully relaxed such that the Hellman-Feynmann forces on the ions are less than 0.001 eV/\AA and the total pressure on the cell is close to zero. Total energy of each of the cases has been computed and compared. The electronic density of states and the band structure were computed on the lowest energy structure. Polarization within the insulating state of the system has been computed by Berry phase method \cite{Vanderbilt}. The result was further corroborated by the polarization obtained from Born effective charges computed using density functional perturbation theory.  

\section{Results and Discussion}
Figure 1 shows the typical remanent hysteresis loops of the sample measured at different temperatures and also the variation of the remanent polarization with temperature. The electronic ferroelectric polarization $P_{el}$ (discussed later), estimated from the x-ray diffraction data, are also shown. The measurement of remanent hysteresis loops employs a specially designed protocol which eliminates the contribution from non-remanent and non-hysteretic polarization components \cite{Chowdhury-2}. This protocol consists of sending out a train of fourteen voltage pulses which measure the hysteresis loops formed from the contribution of remanent, nonremanent, hysteretic and nonhysteretic polarizations as well as from the nonremanent and nonhysteretic polarizations only. Subtraction of the latter loop from the former one yields the intrinsic remanent hysteresis loop. The salient features of the measurement protocol including the voltage pulse train sent out for measuring the remanent polarization are given in the supplementary document \cite{Supplementary}. The observation of small yet finite remanent ferroelectric polarization ($\sim$0.01 $\mu$C/cm$^2$) ensures emergence of ferroelectricity at $T_N$. The evolution of time scale along the hysteresis loop is counterclockwise which is consistent with true ferroelectric behavior and rules out charge injection. It is important to note that application of this specific protocol on several compounds, either improper ferroelectrics with tiny remanent polarization or nonferroelectrics with no remanent polarization, is found to be effective in extracting the characteristic $P-E$ loop to determine the magnitude of remanent polarization \cite{Chowdhury-2}. A rather small nonlinearity in the left and right arms of the $P-E$ loop could possibly manifest the role of ferroelastic switching as well. The loops have been blown-up in the supplementary document to show the extent of nonlinearity of these side arms clearly \cite{Supplementary}. Of course, in general, the nonlinearity in the side arms is expected to be small in remanent hysteresis loops \cite{Supplementary}. Square-looking ferroelectric hysteresis loops have previously been observed in a few cases such as in electrets, in orthoferrite SmFeO$_3$ as well as in thin films of PbTiO$_3$ of thickness 129 nm grown on 0.7 wt\% Nb-doped SrTiO$_3$ with Pt top electrode. Electrets exhibit polarization which diminishes with time \cite{Reaster}. This is not the case with our samples. In a recent work \cite{Podgorny}, quantitative analysis of the hysteresis loop shape using dielectric portraits is shown to offer more accurate information about the thickness of ferroelectric dead layer and its nature - Schottky barrier type or other - and, therefore, may have, at least, peripheral relevance to the loop shape observed here. The observations ealier made for orthoferrites such as SmFeO$_3$ are attributed to improper polarization, believed to originate from exchange striction giving rise to polar displacement of the oxygen ions at the magnetic domain walls \cite{Bellaiche}. On the other hand, observations of similar square loops in PbTiO$_3$ thin films as well as in Pb(Mg$_{1/3}$Nb$_{2/3}$)O$_3$-PbTiO$_3$ composite (typical results shown in the supplementary document \cite{Supplementary}) originate from the switching of 90$^o$ domains \cite{ref} at higher frequency instead of complete 180$^o$ domain reversal which could possibly require lower frequencies. In the present case, of course, we observe complete 180$^o$ reversal of domains as witnessed by piezoresponse force microscopy (PFM), as discussed in the next paragraph. We also observe nearly frequency independent remanent hysteresis loops \cite{Chowdhury-2}.
  
The temperature dependence of remanent polarization suggests a sharp drop in the polarization in the vicinity of $T_N$ ($\sim$600 K) which is indicative of coupling between magnetic and electrical ordering and perhaps a structural transition at this temperature which we further explore using temperature dependent x-ray and Raman studies. We conducted piezoresponse force microscopy (PFM) to explore the ferroelectric switching in the samples. PFM was used in a spectroscopic mode in which a dc bias voltage is applied in a cyclic manner with tip remaining fixed. This yields a local piezoelectric loop which is basically the manifestation of the local piezoelectric vibration on the voltage sweep. To observe the polarization switching, a sequence of dc voltage in a triangular sawtooth form was applied with simultaneous application of 2 V ac voltage to record the corresponding piezoresponse, measured during the ''off'' state at each step to minimize the effect of electrostatic interactions, resulting in a phase-voltage hysteresis loop. PFM amplitude and phase images acquired in PFM dual ac resonance tracking imaging mode, using a cantilever of stiffness 2 N/m and Ti/Ir tip. Figure 2 shows the amplitude and phase contrast PFM images recorded under +100V and -100V dc bias. Two types of sub-micron-sized domains with dark purple and white colors could be seen in the phase contrast image captured under -100V. These are antiparallel domains, also called 180$^o$ domains, where polarization vector is oriented in phase with the applied voltage for purple and out of phase for white. The orientation changes upon switching the electric field. In principle, orthorhombic structure can also exhibit 60$^o$, 90$^o$, and 120$^o$ domains \cite{Damjanovic} and presence of multiple colors indeed points toward existence of these domains, albeit, in small proportions. The 180$^o$ domains, of course, are the dominant ones. The complete switching spectroscopy PFM was also carried out and the strain and phase switching angle versus field loops are shown. The 180$^o$ switching of the domains under $\pm$100 V bias also indicates that during the measurement of remanent hysteresis loop saturation of polarization is achieved as identical bias voltage was applied in that case too. The butterfly shape of the strain versus electric field loop is indicative of the presence of piezoelectric activity in the sample. The distortion of the strain-field loop possibly originates from difference in electrode-sample interface charge structure between top and bottom electrodes. It is important to point out here that though the PFM measurements have been carried out on polycrystalline samples where the conductivity might have finite variation across the grain-grain boundary network, the images recorded indeed show the ferroelectric domains and their switching. The comparison of the topological and phase-contrast PFM images shows that the pattern observed in the topological image is quite different from the pattern observed in the phase-contrast PFM image. Moreover, we have carried out the measurements at different places of the sample to ensure presence of finite intrinsic piezoresponse in the sample. Therefore, influence of conductivity fluctuation on the PFM images is ruled out. It is further mentioned that since the PFM data have been recorded on a polycrystalline sample, one does not know the orientation of individual grains. Therefore, it is not possible to identify the crystallographic direction or plane of the measurement. While polarization vector could be oriented along a-axis (described later), if the grains have orientations that are not perpendicular to the a-axis, one would still observe the ferroelectric switching. Hence, since one serves the switching, one is witnessing the contribution of the component of the polarization along the direction of the applied bias field under the PFM tip from a randomly oriented grain.  

\begin{figure}[!ht]
  \begin{center}
    \subfigure[]{\includegraphics[scale=0.25]{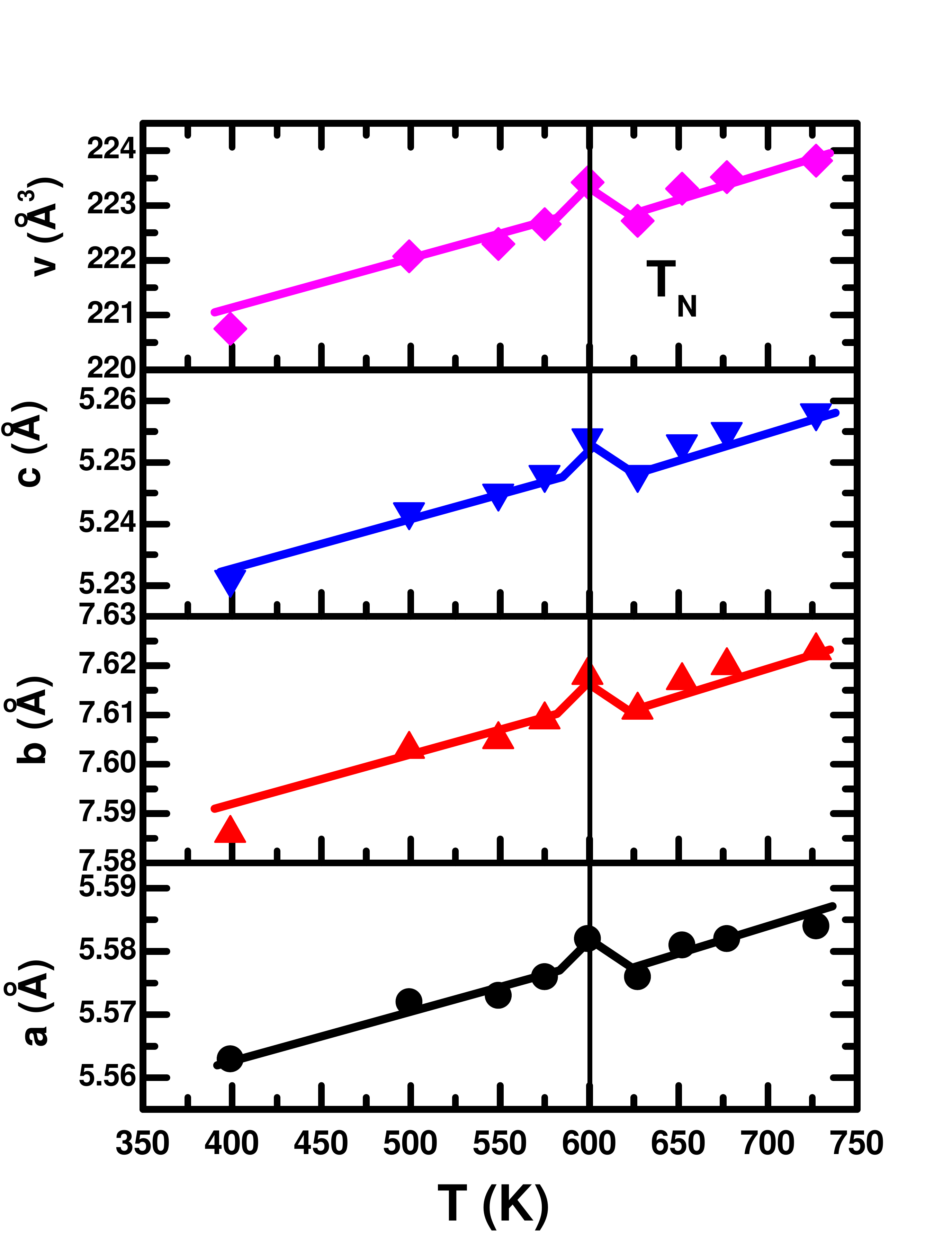}} 
   \subfigure[]{\includegraphics[scale=0.25]{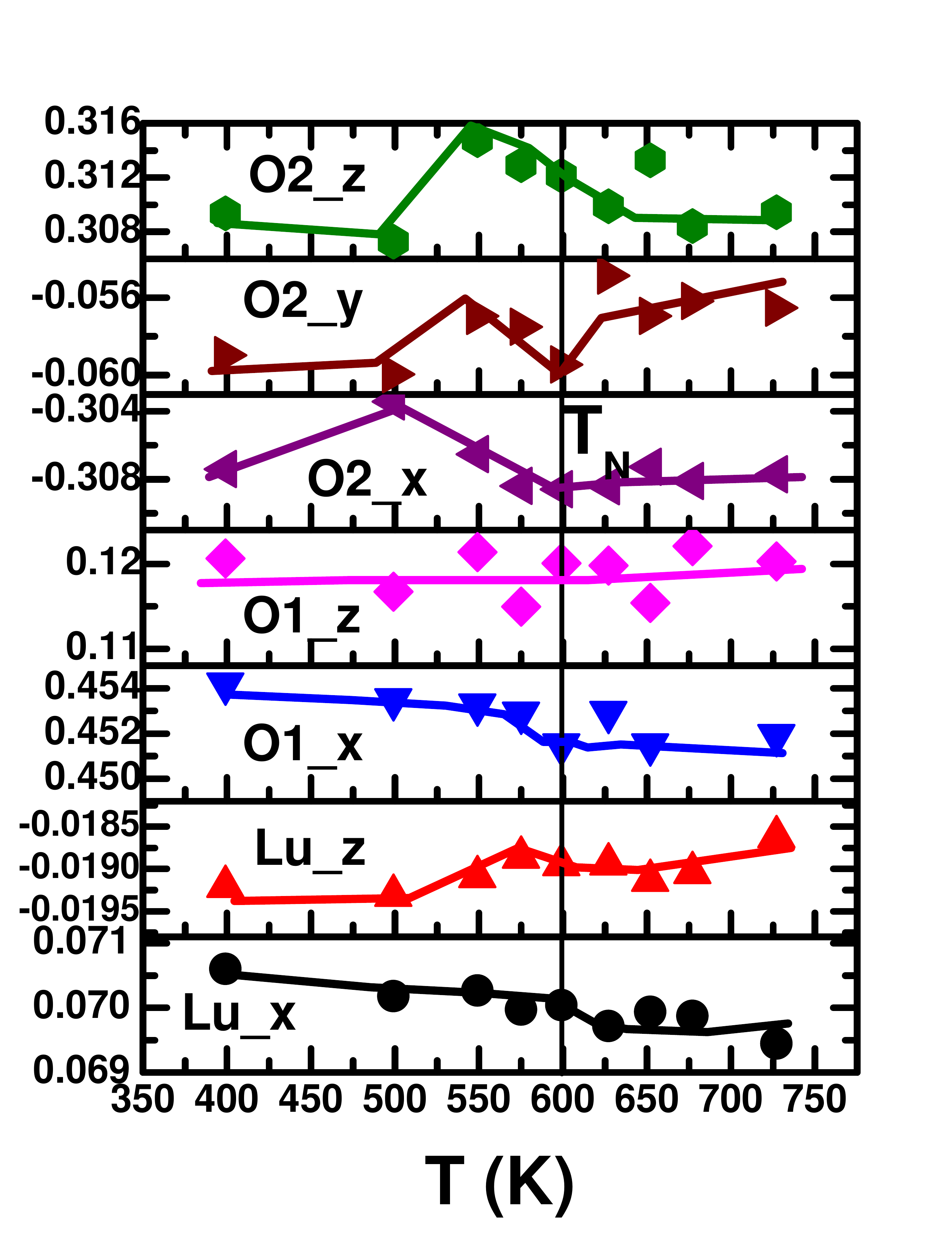}}
    \end{center}
  \caption{(color online) Variation of the (a) lattice parameters and (b) ion positions with temperature obtained from Rietveld refinement of the x-ray diffraction data. }
\end{figure}

\begin{figure}[!ht]
  \begin{center}
    \includegraphics[scale=0.35]{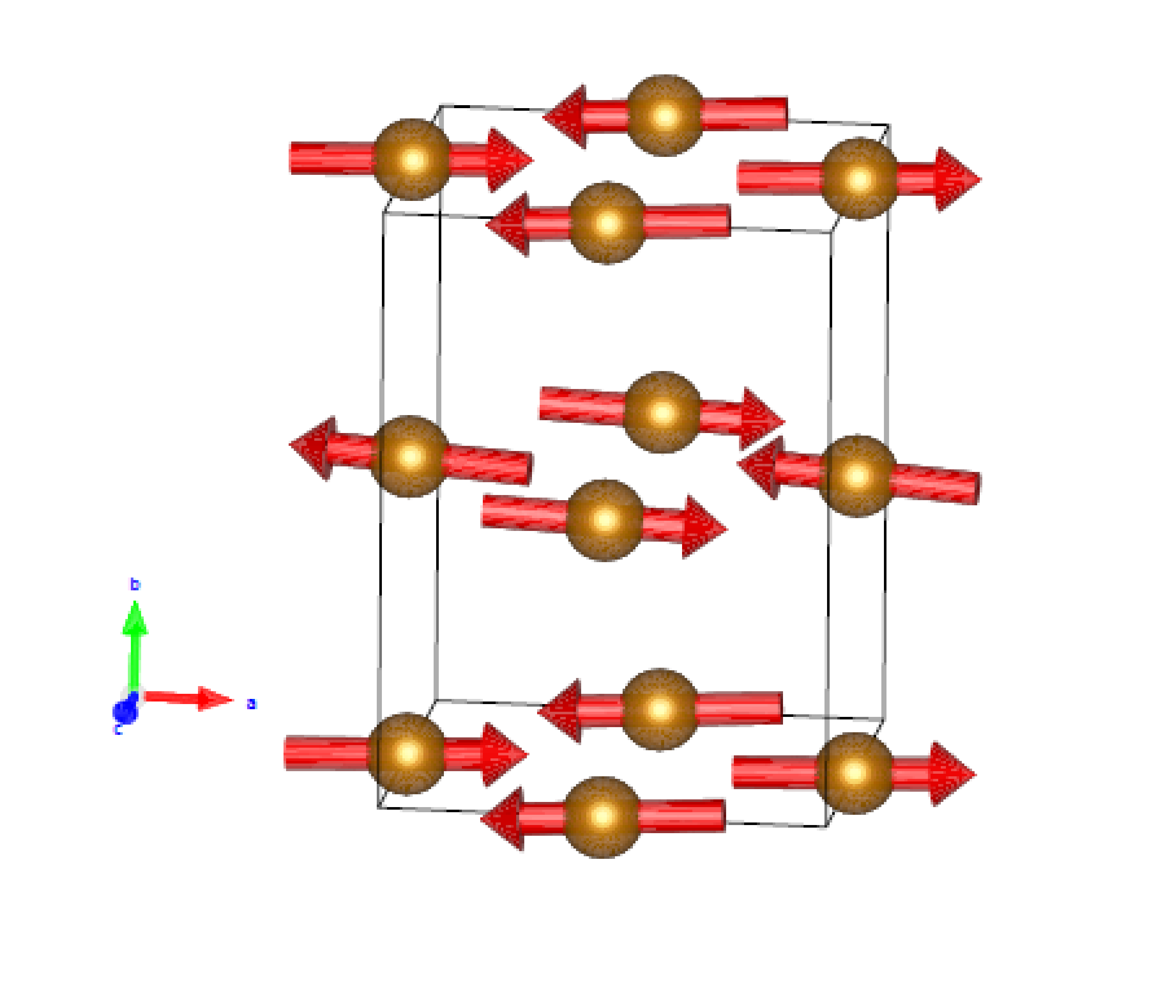} 
    \end{center}
  \caption{(color online) The spin structure of LuFeO$_3$ as per irrep $\Gamma_2$. For clarity, Lu and O ions are not shown.}
\end{figure}

\begin{figure}[ht!]
\begin{center}
    \includegraphics[scale=0.30]{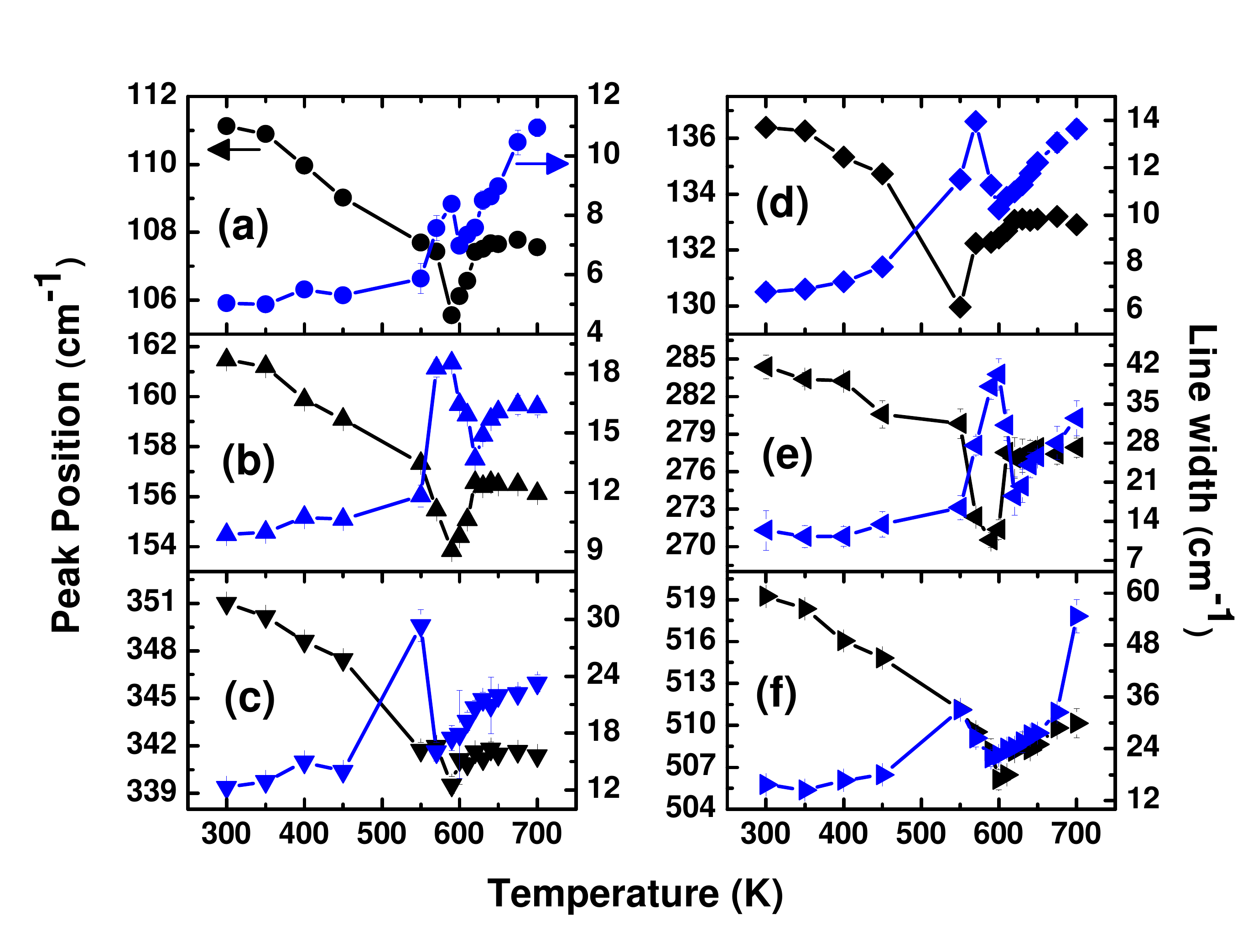} 
    \end{center}
\caption{(color online) The temperature dependence of Raman mode frequency and linewidth. }
\end{figure}

Using the synchrotron x-ray and neutron data, we now examine the contribution of lattice and electronic structure to the overall polarization. Within the limit of the resolution of diffraction data and accuracy of the Rietveld refinement, it appears (reliability factors and goodness of fit vary within 1.0\%-2.5\% \cite{Supplementary}) that the crystallographic structure of the sample remains nonpolar orthorhombic $Pnma$ throughout the entire temperature range. Therefore, if at all there is any structural transition around $T_N$, it is of isostructural type. The isostructural transition is rare and has implications for phonon dynamics. The phonon symmetry does not change across the transition. The calorimetric trace and dc resistivity versus temperature measurements reveal finite latent heat and resistivity hysteresis associated with the transition. The results are included in the supplementary document \cite{Supplementary}.  Although the physics behind isostructural transition is not quite well understood, there are suggestions that this could be due to interaction of electrons with lattice vibrations \cite{Barma}. Both the x-ray and neutron scattering offer evidence of prevalence of nonpolar $Pnma$ structure even below $T_N$. Of course, it is quite possible that extremely small noncentrosymmetry (of the order $\sim$0.16 m$\AA$ mentioned in Ref. 9), if present, remains undetected in these measurements. The small nonlinearity of the left and right sides of the remanent hysteresis loops together with PFM data also suggest a structural transition from nonpolar $Pnma$ to nonpolar yet ferroelastic $P2_12_12_1$ at $T_N$. In Fig. 3, we show the variation of lattice parameters, volume, ion positions etc, determined from the refinement of x-ray data, as a function of temperature across 400-727 K. The estimated standard deviation, obtained during refinement, varies within 0.2\%-0.4\% for all the parameters. It represents the corresponding error bar. Clear anomaly could be observed in almost all the parameters around $T_N$ signifying presence of strong spin-lattice coupling. The nature of the anomaly is similar in all the three lattice parameters and hence the volume; they exhibit anomalous expansion at the onset of magnetic order at $T_N$. Temperature dependent evolution of the lattice parameters, bond lengths/angles together with in- and out-of-phase octahedral tilt and A-site ion displacement has earlier been tracked \cite{Lightfoot} across 25-1285 K for other members of the rare-earth orthoferrite family with large tolerance factor such as LaFeO$_3$. Studies have also been done on orthorhombic PrFeO$_3$ and NdFeO$_3$ \cite{Sosnowska} and on R$_{0.5}$R$_{0.5}$'FeO$_3$ (R = Sm, R' = Pr, Nd) \cite{Buryy} and evidence of spin-lattice coupling could be gathered from anomalies around $T_N$ (650-750 K). In the present case, the Lu, O1, and O2 ions exhibit anomalous displacement below $T_N$; the position of Fe ion is fixed at (0,0,0.5). Within the limit of accuracy with which the positions of the ions have been determined (error bar varies within 0.2\%-0.4\%), it appears that the anomalous displacement of Lu, O1, and O2 ions is consistent with the irreducible representation $\tau_1$ \cite{Lee-2}. This signifies occurrence of isostructural transition at $T_N$. The allowed irreducible representations corresponding to the anomalous ion displacements at $T_N$, obtained from the group theoretical analysis, as well as the basis vectors for the $\tau_1$ mode are given in the supplementary document \cite{Supplementary}. The comparison of crystallographic parameters determined from x-ray and neutron diffraction \cite{Supplementary} shows that though there exists some difference between the numerical values of the parameters the overall trend is quite similar. It is important to point out here that because of poorer scatteing of x-ray by the lighter ions such as oxygen, it is difficult to determine the position of oxygen ions accurately from x-ray diffraction data. On the other hand, for commensurate magnetic structure with $\textbf{k}$ = 0, determination of ion positions from neutron diffraction poses problem as both the magnetic and nuclear peaks appear at the same point in reciprocal lattice space. In this case, it is necessary to collect the neutron data at a spallation source across much larger $Q$ range in order to eliminate the influence of magnetic peaks. The consistency in the structural parameters determined from both x-ray and neutron data reflects the accuracy of the results obtained for the present case. The commensurate $\textbf{k}$ = 0 magnetic lattice determined from the neutron diffraction experiments is found to be collinear (Fig. 4), which corroborates the observations made in orthorhombic SmFeO$_3$ \cite{Kuo,Johnson}. It is found, however, that the magnetic lattice for LuFeO$_3$ across 400-700 K could be described by the single irrep $\Gamma_2$ \cite{Supplementary}; corresponding spin configuration is $F_xC_yG_z$ (Fig. 4). This is consistent with nonpolar structure. Below $\sim$400 K, spin-reorientation transition could be observed. We, of course, concentrate here on the data across 400-700 K as we are concerned about ferroelectricity right below $T_N$.

\begin{figure}[ht!]
 \begin{center}
    \includegraphics[scale=0.30]{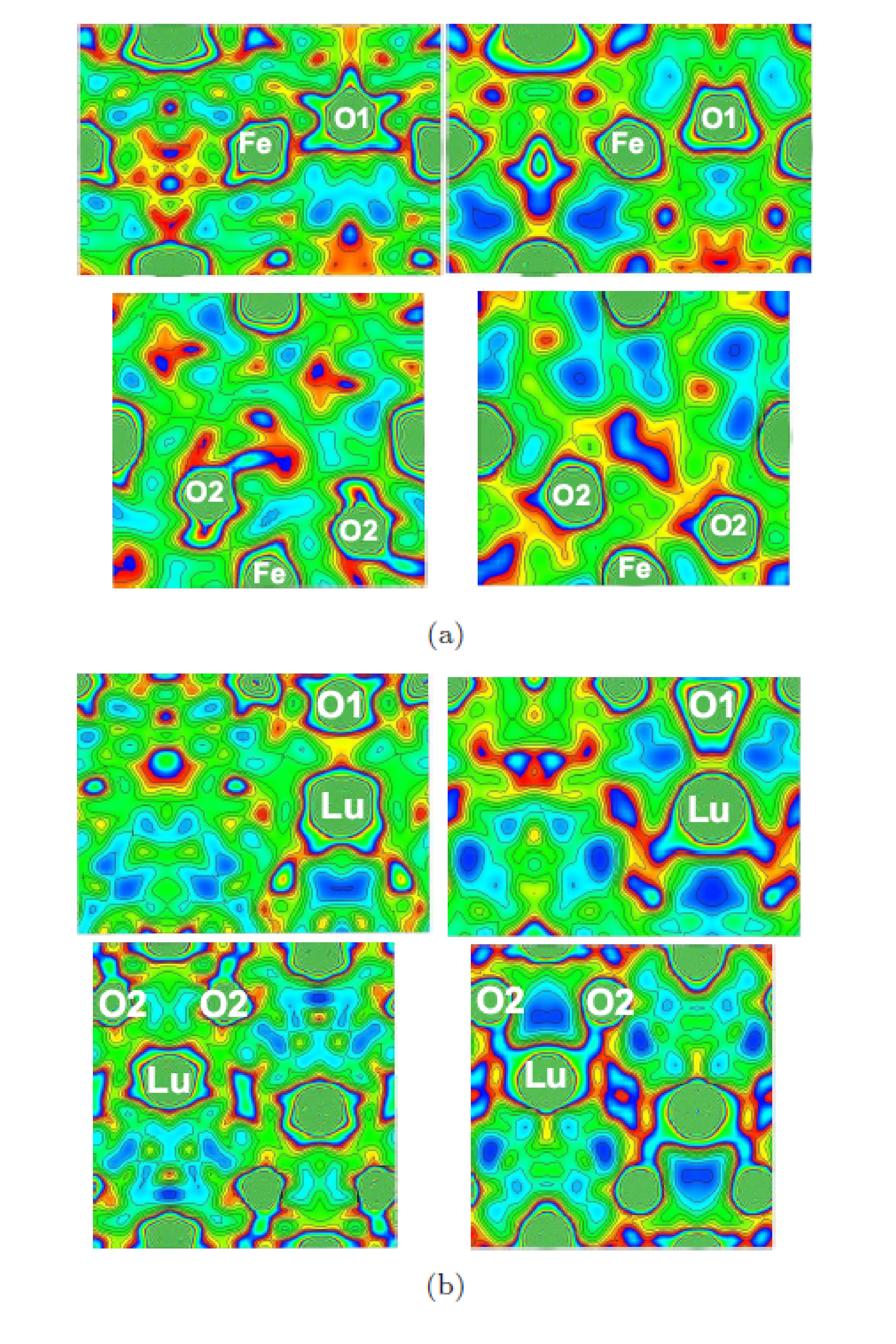} 
    \end{center}
\caption{(color online) The two-dimensional map of electron charge density in (a) (100) and (010) planes showing Fe-O1 and Fe-O2 bonds respectively and (b) (30$\bar{1}$) and (10$\bar{1}$) planes showing Lu-O1 and Lu-O2 bonds respectively; the left panels show the data at 399 K (below $T_N$) while the right ones show the data at 727 K (above $T_N$).}
\end{figure}

\begin{figure}[ht!]
\begin{center}
   \subfigure[]{\includegraphics[scale=0.25]{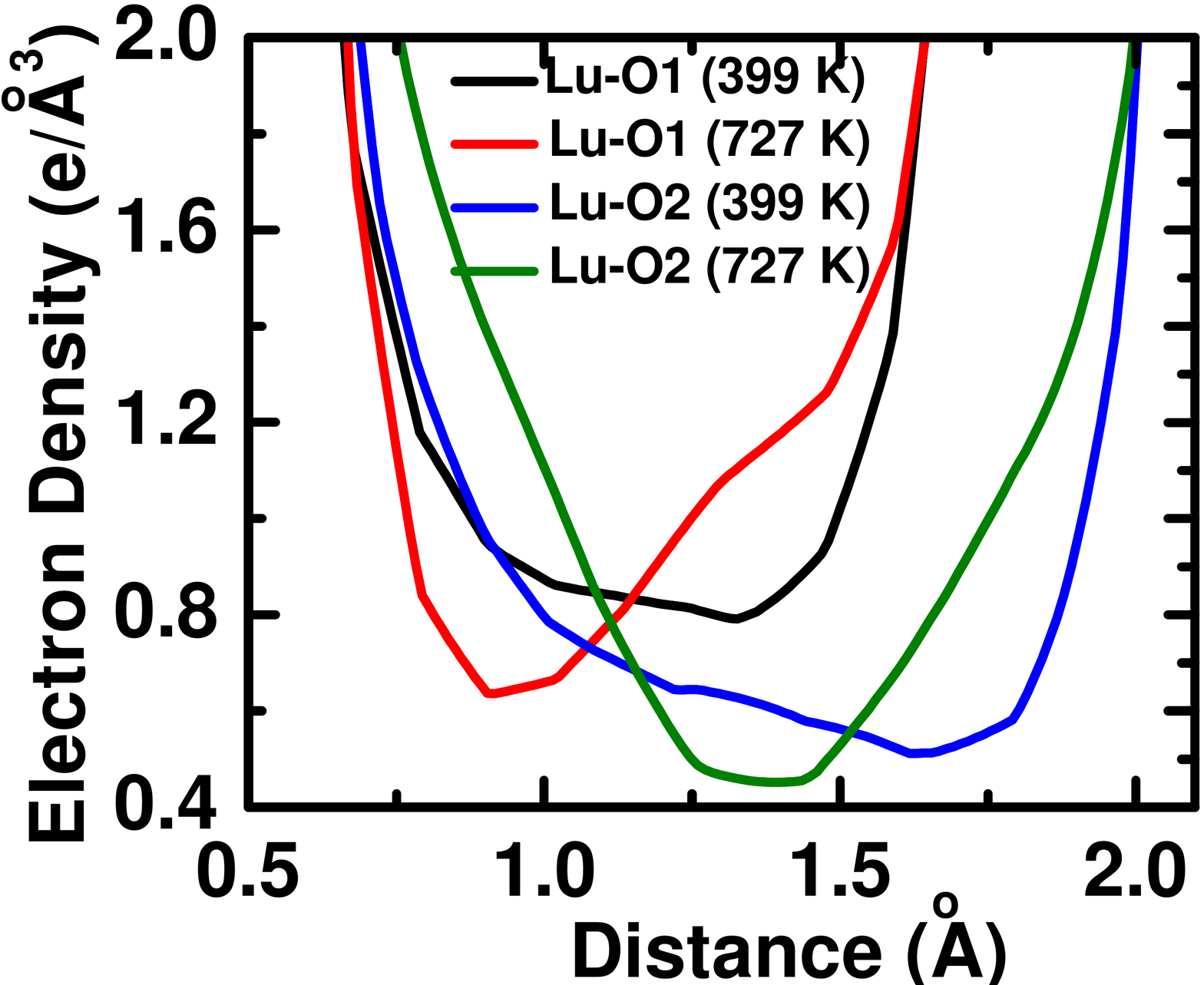}} 
   \subfigure[]{\includegraphics[scale=0.25]{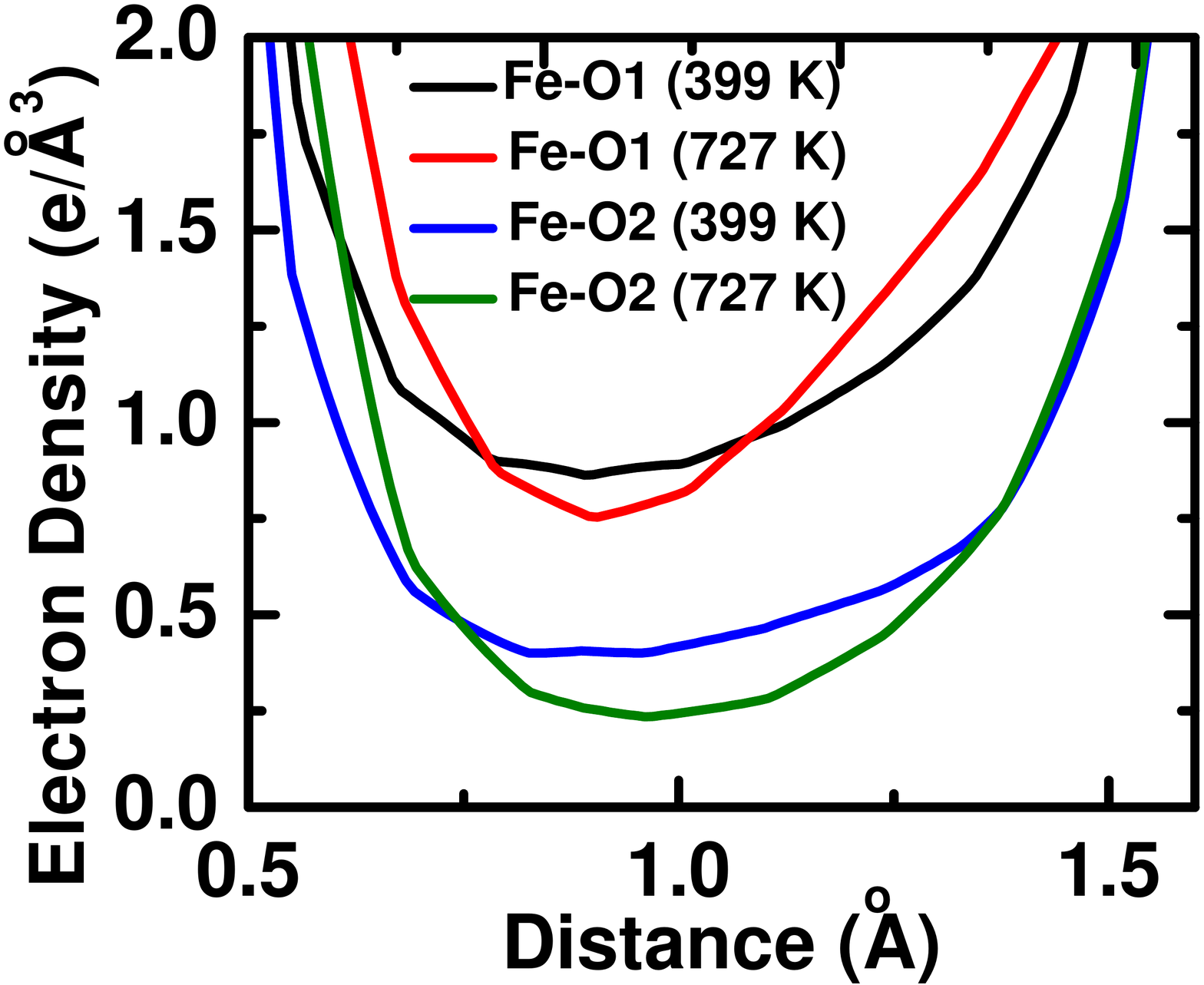}}
   \end{center}
\caption{(color online) The one-dimensional map of variation of electron density across (a) Lu-O and (b) Fe-O bonds at 399 and 727 K. }
\end{figure}

We further investigate the role of lattice by carrying out Raman spectrometry across 300-700 K within the Raman shift range 90 to 1000 cm$^{-1}$ \cite{Supplementary}. The $A_g$ and $B_{1g}$ modes \cite{Venugopalan} could be seen and their frequency shift and linewidth are shown in Fig. 5 as a function of temperature. Distinct anomaly in both frequency shift and linewidth could be observed at $T_N$. However, unlike the phonon softening observed in systems with displacive ferroelectricity arising out of even orthorhombic $Pnma$ to orthorhombic $P2_1ma$ phase transition \cite{Taniguchi}, the phonon modes here exhibit a pattern expected in isosymmetric phase transition where the phonon modes are symmetric both at above and below the transition \cite{Scott-1} and the frequency changes only slightly (Cochran's exponent $<$ 0.1). Though reports exist \cite{Scott-2} for other compounds where deviation from the expected mode softening features could be observed even when the lattice ferroelectricity is finite, in the present case, we could not observe similar feature. Therefore, at best, clear evidence of lattice ferroelectricity appears to be undetectable in the experiments. The theoretical calculations (described later) show that the nonpolar to polar phase transition at $T_N$ could remain undetectable experimentally because of very small energy difference between these phases at below $T_N$.

We then determined the electron charge density distribution over the unit cell. Application of MEM/Rietveld refinement to the high energy synchrotron x-ray diffraction data yields the charge density distribution. The accuracy of MEM in determining the charge density distribution and thus covalency of the bonds has already been established for different compounds including compounds containing combination of heavy elements such as Pb together with light elements such as O \cite{Kuroiwa}. The MEM has also been used to observe the Mn3$d_{x^2 - y^2}$ orbital order \cite{Takata}. In the present case, MEM analysis has been carried out by dividing the unit cell into 48$\times$72$\times$48 pixels for all the temperatures. The details of the refinement and fit statistics are given in the supplementary document \cite{Supplementary}. In Fig. 6, we show the two-dimensional maps of charge density distribution in (100), (010), (30$\bar{1}$), and (10$\bar{1}$) planes at 399 (i.e., below $T_N$) and 727 K (i.e., above $T_N$) in order to show the charge density across Fe-O1, Fe-O2, Lu-O1, and Lu-O2 bonds, respectively. The background charge density is $\sim$0.2 e/\AA$^3$ and the contours lines are mapped across 0 to 1 e/\AA$^3$ at an interval of 0.1 e/\AA$^3$. The charge density around the midpoints of Lu-O1, Lu-O2, Fe-O1, and Fe-O2 bonds at 727 K are 0.636 e/\AA$^3$, 0.452 e/\AA$^3$, 0.754 e/\AA$^3$, 0.235 e/\AA$^3$, respectively. The corresponding figures at 399 K are 0.792 e/\AA$^3$, 0.512 e/\AA$^3$, 0.863 e/\AA$^3$, and 0.401 e/\AA$^3$, respectively. The covalency of all the Lu-O1, Lu-O2, Fe-O1, and Fe-O2 bonds has increased below $T_N$ by different extent. The one-dimensional map of charge density distribution across Lu-O and Fe-O bonds both at 399 and 727 K are shown in Fig. 7. Different extent of charge transfer below $T_N$ leads to asymmetric distribution of charges. The electrons are counted within the minimum charge density surface around each ion of the cell. The difference between the electron count and atomic number gives the charge state ($n$) of the ion. The charge states for Lu, Fe, and O ions turn out to be +3.93, +3.96, -3.53 (for four O1 ions), and -2.20 (for eight O2 ions) repectively at 727 K. At 399 K, the corresponding figures for Lu and Fe ions are +3.80 and +3.50, respectively; O1 ions appear to be of -3.20 charge state while O2 are of -2.0. It appears then that charge disproportionation by nearly 10\%-12\% has taken place among Lu, Fe, and O ions as a result of magnetic transition. Comparable extent of charge disproportionation has earlier been observed \cite{Wright} in Fe$_3$O$_4$ below its charge order (Verwey) transition ($T_{CO}$ $\sim$120 K). In order to calculate the ferroelectric polarization below $T_N$, if any, as a result of off-centric charge density distribution, we first find out the center of charge density distribution contour ($c$) for each ion \cite{Supplementary}. Using this result for all the cations and anions of the unit cell, the net off-centred shift ($\Delta c$) in the charge density distribution contours or charge cloud has been determined. Remarkably, $\Delta c$ turns out to be finite, though small ($\approx$0.003-0.004 \AA), below the $T_N$ and using the relation $P_{el}$ = $n.e.\Delta c/V$ where $e$ = charge on an electron and $V$ = volume of unit cell, we determine the ferroelectric polarization resulting from off-centred charge density distribution within a unit cell ($P_{el}$). Interestingly, as against the observation made earlier \cite{Lee-1}, the $P_{el}$ turns out to be oriented along a-axis. We plot the values of $P_{el}$ as a function of temperature in Fig. 1. The order of the magnitude of $P_{el}$ appears to be comparable to what has been found from direct measurement (of the order of $\sim$0.01 $\mu$C/cm$^2$ at 300 K). This result indicates that the tiny ferroelectric polarization measured in orthorhombic LuFeO$_3$ could possibily have electronic origin. In spite of lattice centrosymmetry, consistent with $\Gamma_2$ irrep, redistribution of charges below $T_N$ could yield a finite global ferroelectric polarization. In LaMn$_3$Cr$_4$O$_{12}$ too, finite electronic ferroelectricity was claimed to result from collinear spin ordering \cite{Wang} within a cubic lattice. Of course, as pointed out earlier, a theoretical work \cite{Bellaiche} has recently suggested that polar displacement of ions at the antiferromagnetic domain walls could induce finite ferroelectric polarization even at $T_N$ in orthorhombic SmFeO$_3$. This is proposed to be true for other rare-earth orthoferrites as well.   

\begin{figure*}[!ht]
 \begin{center}
    \includegraphics[scale=0.50]{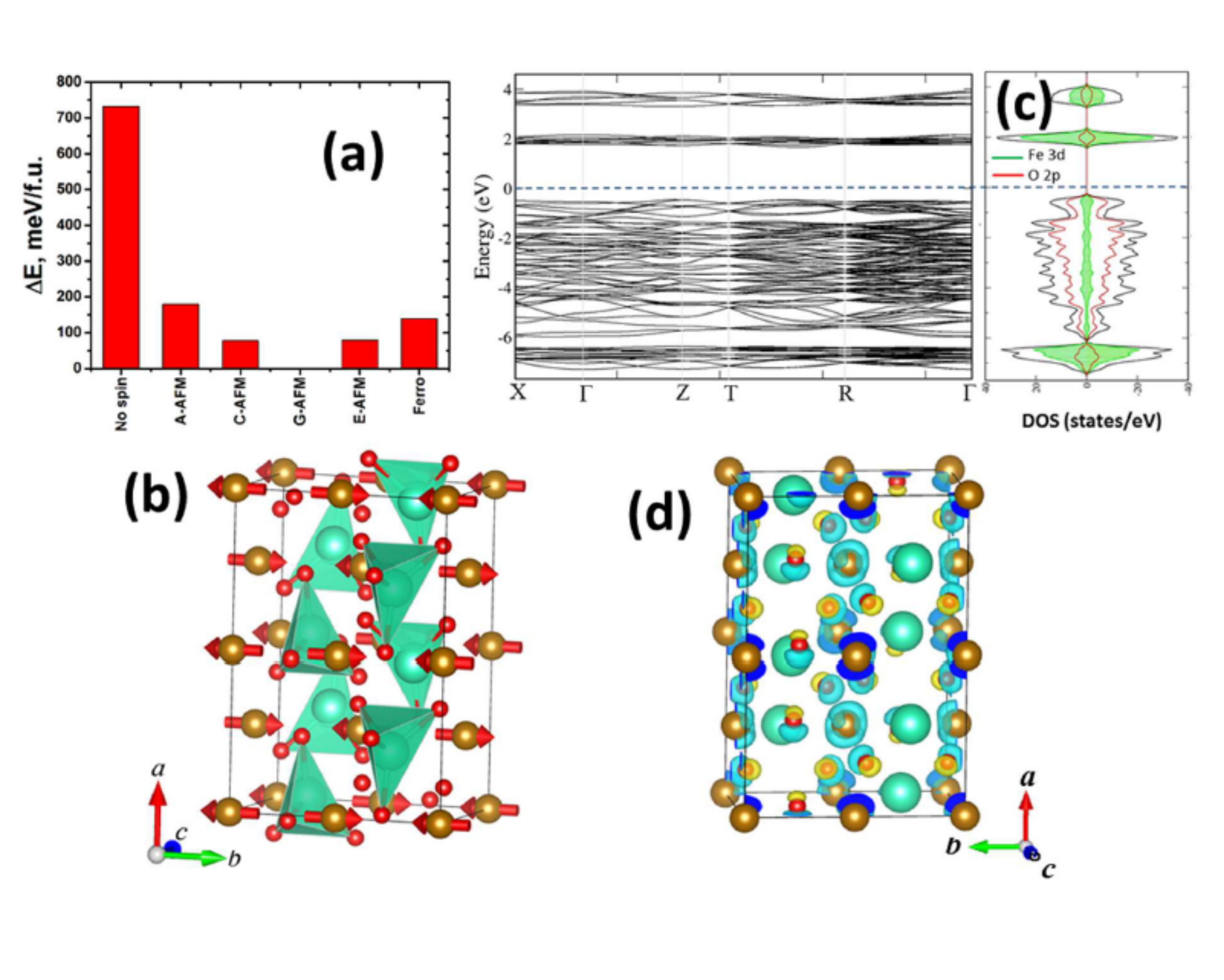} 
    \end{center}
  \caption{(color online) (a) Comparison of total energy of LuFeO$_3$ for different spin configurations; (b) schematic of LuFeO$_3$ supercell wherein G-AFM is the most favoured spin structure; (c) Electronic band structure and density of states of G-AFM structure; and (d) differential charge density plot of distorted $Pnma$ structure in presence of G-AFM order with and without Hubbard parameter $U_{eff}$. }
\end{figure*}        

Interestingly, we also observe finite piezostriction in orthorhombic LuFeO$_3$. Application of electric field induces detectable piezostriction (Fig. 2) as a result of reasonably large dielectric constant \cite{Chowdhury-1} and electrostrictive coefficient \cite{Li}. The ferroelectric domains observed in PFM actually represent those for the lattice. How they are related to the electronic ferroelectric domains, if present, is not yet understood. Of course, as pointed out above, presence of subtle noncentrosymmetry below $T_N$ cannot be completely ruled out and it needs further investigation.  Observation of lattice noncentrosymmetry only under an electric field in presence of electronic ferroelectricity has earlier been noted in a charge-transfer complex tetrathiafulvalene-chloranil \cite{Kobayashi}. It will be interesting to search for similar result in other purely electronic ferroelectric systems.

Since the origin of ferroelectricity in orthorhombic LuFeO$_3$ is not quite conclusively understood, we employed first-principles density functional theory (DFT) based calculations to investigate two distinct possibilities: (i) long-range magnetic order mediated hybridization of electronic oribitals leading to the asymmetric charge density distribution and hence finite electronic ferroelectricity and (ii) breaking of spatial inversion symmetry of the crystallographic structure in presence of magnetic order. The third possibility of finite ferroelectric polarization below $T_N$ as a result of exchange-striction driven lattice ferroelectricity at the magnetic domain boundaries has earlier been explored by others \cite{Bellaiche}. In fact, this possibility has started gaining ground in the context of rare-earth orthoferrites as it shows that a single-domain bulk sample cannot support ferroelectricity which is consistent with the experimental data \cite{Kuo}. However, in a real multidomain sample, ferroelectricity emerges at the antiferromagnetic domain boundary. The magnitude of polarization observed experimentally is also consistent with the theoretical prediction. The relevance of this mechanism in the context of describing ferroelectricity in different magnetic ferroelectric systems has been highlighted by Scott and Gardner \cite{ref}. In a separate theoretical work \cite{Chen}, it has also been pointed out that orthoferrites (RFeO$_3$) or orthochromites (RCrO$_3$) with two magnetic sublattices R and Fe/Cr, could exhibit ferroelectric polarization because of non-relativistic exchange striction with large magnetostructural effect. This mechanism could explain the ferroelectricity in DyFeO$_3$ below $T_N^{Dy}$ under a magnetic field \cite{Tokunaga}. Interestingly, a morphotropic mixture of hexagonal and orthorhombic LuFeO$_3$ too, has recently been synthesized in thin film form and ferroelectric polarization along with magnetoelectric coupling have been investigated \cite{Song}.  

\begin{table}

\caption{Calculated structural parameters, magnetic moment, and electronic band gap within $Pnma$ structure.} 

\begin{tabular}{p{0.7in}p{0.5in}p{0.5in}p{0.5in}p{0.5in}p{0.5in}} \hline\hline
 &  GGA+U  & with & PBE &  & GGA \newline with PW\\ 
  &   $U_{eff}$ = & & & & $U_{eff}$ =\\ \hline
  &  0 eV & 3.0 eV & 4.0 eV & 5.0 eV & 0 eV\\ \hline 
$a$ (\AA) &  5.480 & 5.501 & 5.504 & 5.506 & 5.555\\
$b$ (\AA) &  7.472 & 7.501 & 7.504 & 7.508 & 7.574\\
$c$ (\AA) &  5.152 &  5.172 & 5.175 & 5.177 & 5.223\\
$\mu_{Fe}$ ($\mu_B$) & 3.63 & 4.02 & 4.09 & 4.15  & 3.66\\
$E_g$ (eV) & 0.4 & 1.66 & 2.05 & 2.40 & 0.35\\ \hline\hline

\end{tabular}

\end{table}

Figure 8(a) plots the total energy per formula unit in a 2$\times$1$\times$1 supercell relative to the lowest energy spin configuration. It is found that the G-type antiferromagnetic order of the Fe ions corresponds to the lowest energy within the orthorhombic $Pnma$ symmetry. Experimentally determined spin-ordering also conforms to the above calculation. The closest competing spin structure is C-AFM which has $\sim$78 meV/f.u. higher energy over the G-AFM structure. The other spin structures correspond to still higher energies. Therefore, it is concluded that the G-AFM is the most favored spin structure within the magnetically ordered structure of orthorhombic LuFeO$_3$ and all the calculations were performed assuming the above structure of LuFeO$_3$. To study the structural stability of the centrosymmetric $Pnma$ phase of LuFeO$_3$, especially in presence of G-AFM order, as well as to understand the evolution of ferroelectric polarization within an apparent centrosymmetric phase, we performed first-principles density functional theory based calculations. Since G-AFM ordering of the Fe ions has been found to be the most favored magnetic ordering, we used GGA and GGA+U methods to relax the experimentally obtained structure within G-AFM ordering. Table-I lists the optimized structural details and electronic band-gap of the compound within $Pnma$ symmetry obtained from using different functionals and different $U_{eff}$.  

It is observed that the optimized lattice parameters, obtained from GGA and GGA+U calculations, are underestimation of the lattice parameters obtained experimentally at 298 K ($a$ = 5.574 \AA, $b$ = 7.600 \AA, and $c$ = 5.241 \AA). Such underestimation is not unusual considering the large temperature difference between the structure obtained experimentally (at 298 K) and the one obtained theoretically (at 0 K). In order to avoid bias toward ferroelectric instability, if any, we, however, used the optimized structure (instead of experimentally observed structure) for the calculation of the ferroelectric polarization. Figure 8(b) schematically shows the G-type antiferromagnetic ordering of the Fe-ions in the 2$\times$1$\times$1 supercell of LuFeO$_3$. It is interesting to note that the arrangement of the Lu ions along a-axis is chiral. The electronic band gap within the G-type antiferromagnetic ordering for GGA is estimated to be $\sim$0.4 eV, much smaller than the experimental observation. Such underestimation of band gap by DFT is well known, in particular for strongly correlated systems. We demonstrate that application of DFT+U method is helpful to increase the band gap of the system.  Electronic band structure, total and site-projected density of states with GGA+U ($U_{eff}$ = 4.0 eV), is presented in Fig. 8(c). It is observed that the upper part of the valence band is occupied predominantly by O $2p$ states whereas the lower part of the conduction band is dominated by Fe $3d$ states. The estimated band gap is $\sim$2.05 eV. This is comparable to the observations made ($\sim$2.07 eV) in rare-earth orthoferrites \cite{Kotnana}. Further, the electrical polarization calculations were also done using GGA and GGA+U method with $U_{eff}$ = 4.0 eV. Small variation of $U_{eff}$ was found not to affect the structural stability.

We first explored the possibility of symmetry lowering structural phase transition to a polar group where spontaneous polarization is realized. The calculation of phonon density of states for $Pnma$ structure in presence of G-antiferromagnetic order, of course, confirms the stability of the phonon modes \cite{Supplementary}. However, as suggested in the experiment, one of the subgroups of $Pnma$ is $Pna2_1$ which is one of the possible polymorphs. Any subtle structural phase transition which was not detected by the experimental studies could be further explored by total energy computation using density functional studies. Using the experimental structural parameters within the antiferromagnetic phase at 298 K, we transformed the experimentally determined $Pnma$ structure to $Pna2_1$ using a program, TRANSTRU, within Bilbao crystallographic server \cite{Aroyo}. Comparison of the total energies of the fully relaxed $Pnma$ and $Pna2_1$ structures, shown in Table-II, highlights similar values. The extent of difference in energy between the two polymorphs is smaller than the room temperature thermal energy and therefore could not be distinguished experimentally. Nonetheless, our calculations predict a polar phase ($Pna2_1$) for LuFeO$_3$ in presence of G-type antiferromagnetic spin order although the difference in energy is of the order of temperature fluctuation and, therefore, one cannot be really sure about the phase stability. Since such a small distortion might involve movement of oxygen ions, neutron diffraction experiment at a spallation source is necessary to track the movement accurately. Interestingly, similar calculations performed on relaxed and optimized structures of isostructural yet nonferroelectric LaFeO$_3$ and NdFeO$_3$ show that the centrosymmetric $Pnma$ structure is more stable even in presence of G-AFM order. Given this result, it will be interesting to examine (i) whether rare-earth orthoferrites with tolerance factor ($t$) smaller than a critical value ($t_C$) could exhibit possibility of nonpolar to polar phase transition in presence of magnetic order and (ii) whether engineering of lattice strain could stabilize the polar phase below $T_N$ in epitaxial thin films or nanostructures. They may be addressed in subsequent works.      

\begin{table}

\caption{Comparison of the lattice parameters, magnetic moment, band gap, and difference in total energy between $Pnma$ and $Pna2_1$ structures} 

\begin{tabular}{p{1.0in}p{1.0in}p{1.0in}} \hline\hline
 & $Pnma$ & $Pna2_1$\\ \hline
$a$ (\AA) & 5.480 & 5.480\\
$b$ (\AA) & 7.472 & 5.152\\
$c$ (\AA) & 5.152 & 7.472\\
$\Delta E$ (meV/f.u.)  & 0.00 & -0.23\\
$\mu_{Fe}$ ($\mu_B$) & 3.63 & 3.62\\
$E_g$ (eV) & 0.4 & 0.36\\
\hline \hline

\end{tabular}

\end{table}

We then considered the distorted $Pnma$ in presence of G-AFM structure. Calculation of polarization using modern theory of polarization (Berry phase formalism) predicts a small spontaneous polarization, $\sim$4.6 nC/cm$^2$ at $U_{eff}$ = 0. At $U_{eff}$ = 4.0 eV, the polarization turns out to be $\sim$1.2 nC/cm$^2$. The polarization ($P$) is given by \cite{Saha} \\

$P = Z_\lambda^*u_\lambda = \frac{Z_\lambda^{*2}E}{\omega_\lambda(T)^2} + \frac{\Delta Z_\lambda^*}{\omega_\lambda(T)^2}M(T)$\\

where $Z_\lambda^*$, $\omega_\lambda(T)$, and $u_\lambda$ are the effective charge, frequency of the phonon mode, and the ionic displacement associated with $\lambda$ (an order parameter which assumes the value 0 at the paraelectric phase and 1 at the ferroelectric phase) which, in the present case, is coupled with the spin. The first term is related to the dielectric contribution while the second term arises from the spin-lattice coupling effect. Effect of magnetization vis-a-vis degree of ordering of the spins can be further qualitatively assessed by the comparing the differential charge density plot with and without the application of Hubbard parameter $U_{eff}$, as shown in Fig. 8(d). Figure 8(d) shows that upon application of the $U_{eff}$, the charge distribution over the oxygen ions is modified. Under such condition, it has been found that nearly 30\% of the total polarization $P$ is originated from noncentrosymmetric electronic charge density distribtion while the rest 70\% is from the lattice effect (distorted $Pnma$). Observation of finite electronic ferroelectricity within distorted $Pnma$ structure, in presence of G-AFM order, supports the experimental observation of finite charge disproportionation at $T_N$. However, the contribution of lattice too, turns out to be finite. This subtle contribution from lattice could not be clearly detected experimentally as neutron diffraction at a spallation source needs to be carried out. Of course, it is worth mentioning, in this context, that the orthorhombic distortion enhances by more than 2.5\% below $T_N$ \cite{Supplementary}. This could be the reflection of subtle distortion within the orthorhombic $Pnma$ lattice which eventually contributes to the polarization too. Although, theoretical calculations, in the present case, do reveal contribution of both electronic and lattice structures to the ferroelectricity, in strongly correlated electron systems, decoupling of electronic and lattice structural transition is not rare \cite{Chuang,Tao}. The collinear magnetic structure has earlier been predicted to exhibit finite ferroelectricity at the onset of magnetic order because of exchange striction effect \cite{Tokura}. In many cases of rare-earth orthoferrites, concomitant structural transition to a polar phase could not be observed \cite{Johnson}. Our theoretical results point out that, for LuFeO$_3$, this is due to the comparable energy scales of the polar and nonpolar phases. Of course, in sharp contrast to the observations made in orthorhombic LuFeO$_3$, isostructural yet nonferroelectric LaFeO$_3$ and NdFeO$_3$ do not exhibit electronic and/or lattice ferroelectricity in presence of G-AFM. They also do not exhibit any instability towards $Pna2_1$ phase at the onset of G-AFM order. This result highlights the bias toward ferroelectricity in LuFeO$_3$ in presence of G-AFM. As pointed out earlier, finite ferroelectricity could possibly emerge in rare-earth orthoferrites in presence of magnetic order only if their tolerance factor ($t$) is smaller than a critical value $t_C$. In order to estimate the spontaneous polarization for LuFeO$_3$ using Born effective charges and corroborate the results obtained from Berry phase formalism, we further employed $\Delta P = [P(u) - P(0)] = \frac{1}{\Omega}\int Z_{ij}^*du$, where $Z_{ij}^*$ represents Born effective charge tensor and $u$ is displacement vector of the ions in the ferroelectric with respect to the paraelectric phase. We calculated the Born effective charge tensor using density functional perturbation theory with GGA+U ($U_{eff}$ = 4.0 eV). Table-III lists the principal elements of Born effective charge tensors of the ions within the distorted G-AFM structure. Interestingly, for $U_{eff}$ = 4.0 eV, all the ions - Lu, Fe, and O - exhibit a maximum $\sim$33\% rise in their effective charges - +3, +3, and -2, respectively. Such anomalous change in the effective charges for Fe and O ions is indicative of a sizeable covalent character of Fe-O bonds in LuFeO$_3$. In fact, our electron localization function (ELF) calculation demonstrates asymmetric distribution signifying preferential accumulation of charge at one end of a bond which, in turn, is indicative of covalency. Application of $U_{eff}$ appears to be reducing the effective charge of Fe and O while that of Lu remains nearly the same. The polarization $P$, estimated from the Born effective charges, turns out to be comparable to what has been found from Berry phase formalism. We further mention here that even though $U_{eff}$ influences the band gap and Born effective charge of the ions, no clear $U_{eff}$ dependence of $P$ could be noticed.

\begin{table}

\caption{Elements of Born Effective Charge tensors of ions in presence of G-AFM structure.} 

\begin{tabular}{p{0.3in}p{0.3in}p{0.3in}p{0.5in}p{0.3in}p{0.3in}p{0.3in}} \hline\hline
    &  $U_{eff}$    &  = 0.0 & eV & $U_{eff}$ &  = 4.0 & eV\\
Ion & $Z_{xx}$ & $Z_{yy}$ & $Z_{zz}$ & $Z_{xx}$ & $Z_{yy}$ & $Z_{zz}$\\ \hline
Lu & 4.09 & 3.50 & 3.79 & 3.95 & 3.48 & 3.87\\
Fe & 6.20 & 5.59 & 5.16 & 4.08 & 3.82 & 3.84\\
O1 & -2.46 & -4.06 & -2.85 & -2.11 & -3.05 & -2.54\\
O2 & -3.91 & -2.51 & -3.05 & -2.96 & -2.12 & -2.59\\ \hline \hline

\end{tabular}

\end{table}

It is true that the theoretically calculated polarization ($P$), for distorted $Pnma$ structure in presence of G-AFM and $U_{eff}$ = 4.0 eV, is smaller than that observed experimentally by a factor of nearly six. This could be because the calculations do not take care of the small lattice strain, domain boundaries, defect network etc which could always be present in as-prepared samples unless special care is taken to remove them. The strain-field in a real sample (even in bulk form) could couple with ferroelectric instability and offer higher polarization as a consequence. However, from the theoretical calculations carried out in this work, it is clear that this orthorhombic LuFeO$_3$ compound (even in its strain-free most pristine form), in presence of G-antiferromagnetic order, is an electronic ferroelectric with lattice structure residing very close to the ferroelectric instability. 

\section{Summary}
In summary, we show by using synchrotron x-ray, neutron, piezoresponse force, and remanent hysteresis data that small but finite ferroelectricity indeed emerges below $T_N$ even in the bulk sample of orthorhombic LuFeO$_3$. An earlier work [Phys. Rev. B $\textbf{96}$, 104431 (2017)] showed that this could result from lattice ferroelectricity due to exchange striction at the antiferromagnetic domain boundaries within a multidomain system. The present work highlights that the ferroelectricity may have electronic origin as charge disproportionation takes place and bond covalency enhances below $T_N$. The first-principles calculations show that within the distorted $Pnma$ structure, in presence of G-antiferromagnetic order, ferroelectricity has small but finite contribution from both electronic and lattice structures. Theoretical calculations also highlight the possibility of structural transition from $Pnma$ to $Pna2_1$ at $T_N$. However, because of tiny energy difference between the phases (smaller than the room temperature thermal energy), the structural transition could not be detected experimentally. In presence of G-antiferromagnetic order, this compound, therefore, in pristine and single domain form, is possibly, primarily, an electronic ferroelectric with lattice structure residing very close to ferroelectric instability. Under an electric field, tiny yet detectable piezostriction, of course, could be noticed. Small lattice strain, present even in `as-prepared' bulk sample, could also yield lattice ferroelectricity.    

\begin{center}
$\textbf{ACKNOWLEDGMENTS}$
\end{center}

Two of the authors - UC and SG - acknowledge Research Associateship and Senior Research Associateship of CSIR, respectively. The authors UC, SG, and AR have equal contribution to this work. The authors DB and AG acknowledge support from SERB, Govt. of India (project no. EMR/2016/001472). The author DB also acknowledges a collaborative research program (CRS-M-201) with Bhabha Atomic Research Centre, Mumbai, India. The author AR acknowledges support from SERB, Govt. of India (project no. YSS/2014/000287). The author TC gratefully acknowledges the financial support provided by FRM-II to perform the neutron scattering measurements at the Heinz Meier-Leibnitz Zentrum (MLZ), Garching, Germany.

\end{document}